\documentclass[prd,aps,amsmath,nofootinbib,superscriptaddress,showpacs]
{revtex4}

\usepackage{graphicx}
\usepackage{hyperref}
\usepackage{bm}
\usepackage{epsfig}
\usepackage{amsfonts}
\usepackage{mathrsfs}
\usepackage{amsmath}

\catcode`@=11
\let\savesort=\NAT@sort@cites
\newcommand\nosort[1]{\edef\NAT@cite@list{#1}}
\def\citenosort#1{\let\NAT@sort@cites=\nosort \cite{#1}%
   \let\NAT@sort@cites=\savesort}
\catcode`@=12 
\makeatletter
    
    \newcommand{\Rmnum}[1]{\expandafter\@slowromancap\romannumeral #1@}
  \makeatother

\newcommand{\beq}{\begin{equation}}
\newcommand{\eeq}{\end{equation}}

\newcommand{\bea}{\begin{eqnarray}}
\newcommand{\eea}{\end{eqnarray}}
\newcommand{\pa}{\partial}
\newcommand{\bib}{\bibitem}
\newcommand{\comment}[1]{}

\def\hs{\hspace}

\begin{document}

\title{A New Exponential Gravity}
\author{Qiang Xu}
\email{xuqiangxu@pku.edu.cn}
\affiliation{Department of Physics, and State Key Laboratory of Nuclear Physics and Technology,\\
Peking University, Beijing 100871, P.R. China}

\author{Bin Chen}
\email{bchen01@pku.edu.cn}
\affiliation{Department of Physics, and State Key Laboratory of Nuclear Physics and Technology,\\
Peking University, Beijing 100871, P.R. China}
\affiliation{Center for High Energy Physics, Peking University, Beijing 100871, P.R. China}


\begin{abstract}
We propose a new exponential $f(R)$ gravity model with $f(R)=(R-\lambda c)e^{\lambda(c/R)^n}$ and $n>3, \lambda\geq 1, c>0$ to explain late-time acceleration of the universe. At the high curvature region, the model behaves like the $\Lambda$CDM model. In the asymptotic future, it reaches a stable de-Sitter spacetime. It is a cosmologically viable model and can evade the local gravity constraints easily. This model share many features with other $f(R)$ dark energy models like Hu-Sawicki model and Exponential gravity model. In it the dark energy equation of state is of an oscillating form and can cross phantom divide line $\omega_{de}=-1$. In particular, in the parameter range $3< n\leq 4, \lambda \sim 1$, the model is most distinguishable from other models. For instance, when $n=4$, $\lambda=1$, the dark energy equation of state will cross $-1$  in the earlier future and has a stronger oscillating form than the other models, the dark energy density in asymptotical future is smaller than the one in the high curvature region. This new model can evade the local gravity tests easily when $n>3$ and $\lambda>1$.
\end{abstract}

\noindent 

\maketitle

\section{Introduction}
\label{sec:introduction}

As we know, the standard big-bang cosmology based on radiation and matter dominated epochs can be
well described within the framework of General Relativity\cite{gr1,gr2}. The rapid development
of observational cosmology starting from 1990s shows that the  expansion of our universe in the present
epoch is accelerating. Currently, the most successful model of cosmology that we have is the $\Lambda$CDM model.
It is in well match with a wide variety of modern cosmological observations that have stunned the physicist
community in the last decade. However, the $\Lambda$CDM model is not perfect in many aspects. First of all,
the cosmological constant remains a mystery, can not be explained clearly in any known theory. If the
cosmological constant originates from the vacuum energy in quantum field theory, as many people believe,
its energy scale is too large to be compatible with the observed dark energy density\cite{la}. 
Moreover, the observation indicates that the dark energy equation of the state may cross the phantom divide line $\omega=-1$. This suggests that the cosmological constant\cite{Hannestad:2002ur,Cepa:2004bc,cmb} may not be the only candidate for dark energy. There have been proposed  many dynamical dark energy models to explain cosmic acceleration, ranging from quintessence, phantom, quintom to chaplygin gas models. For the nice reviews on dark energy, see \cite{Li:2011sd,Copeland:2006wr,Cai:2009zp}. In  dynamical dark energy models, one has to introduce at least one dynamical scalar to drive late-time acceleration, similar to the scalar driving the early-time inflation.

An alternative scenario for dark energy is infrared(IR) modified gravity. 
Among many IR modified gravity model, $f(R)$ gravity is of particular interest.   One important feature in $f(R)$ gravity is the intrinsic existence of an extra dynamical scalar degree of freedom, besides the massless graviton. Therefore it is possible  to study both the early-time inflation and late-time acceleration of the universe in the framework of $f(R)$ gravity, without introducing ad hoc scalar fields by hand. More interestingly, the effective equation of state  could be smaller than $-1$ in $f(R)$ dark energy models, indicating the scalar behaves like a phantom in the Jordan frame.  On the other hand, the existence of scalar mode is not always pleasant. The fact that the dynamical scalar field may induce a long-range  fifth force suggests that a viable $f(R)$ gravity should satisfy the stringent constraints of local solar system test.

Since the discovery of late-time acceleration of the universe in 1998, the $f(R)$ theory have been extensively
studied as the simplest modified gravity scenario to drive late-time acceleration. The model with a Lagrangian density $f(R)=R-\alpha /R^n$ ($\alpha > 0, n > 0$) was proposed for dark energy\cite{r-n1,ca0,r-n3,Pi:2009an}. However this model is plagued by  matter instability \cite{rn4,rn5} and difficulty to satisfy local gravity constraints. Later on,
researchers have proposed many viable models, seeing \cite{st,hu,tsu,linder,Elizalde:2011ds,Elizalde:2010ts,Cognola:2007zu,Bamba:2010ws}. The Lagrangian of these models have a common form,
adding a function of $g(R/R_0)$ to the Einstein-Hilbert term $R$. In the high curvature region, when $R\gg R_0$, $g(R/R_0)$ tends to be a constant and the model mimic the $\Lambda$CDM model, and should satisfy the local gravity constraints. However, in the late-time universe, the extra terms $g(R/R_0)$ may play a significant role in the evolution.  According to the dynamics of the theory, the dark energy equation of state could cross the phantom divide line, and tends to be $-1$ in the asymptotic future.  For the nice reviews on f(R) theories, see \cite{Nojiri:2010wj,tsu3}.

In this paper, we propose a different $f(R)$ dark energy model that do not contain a cosmological constant.  In our model:
\beq
f(R)=(R-\lambda c)e^{\lambda(\frac{c}{R})^n}, \label{f}
\eeq
where $\lambda\geq 1,c>0$. The stability condition at the asymptotic future requires that $n> 3$. Among three parameters, c has the same dimension as Ricci scalar, $\lambda$ and n are dimensionless
parameters. Different from the usual $f(R)$ models,  the Lagrangian of our model can not be separated into a $R+g(R)$ form. Obviously, in the high curvature region, the exponential factor tends to be $1$ and the model reduces to the $\Lambda$CDM model. Asymptotically, there is a stable de Sitter vacuum, with a different cosmological constant.
We study the cosmological implications of this model.  We discuss when and at what level it modifies the cosmological predictions,  while evade the local tests of gravity. It turns out that even though when $\lambda \gg 1$ or $n\gg 3$, the model becomes indistinguishable from the $\Lambda$CDM model, the model present distinguishable features from other models when $\lambda=1$ and $3< n\leq 4$. We compare our model with other two well-studied dark energy $f(R)$ models,  Hu-Sawicki model\cite{hu} and Exponential gravity model\cite{Cognola:2007zu,linder,Bamba:2010ws}. We find that all of them share some common qualitative features: crossing phantom divide line $\omega_{de}=-1$,  and dark energy equation of state being of an oscillating form, but they differ in details.

This paper is organized as follows. In Section II, after
briefly reviewing  general $f(R)$ theory, we introduce our model and discuss its cosmological implications. Via numerical analysis, we study the evolution of the FRW universe in our model and investigate the evolutions of the dark energy density and dark energy equation of state. In Section III, we analyze the local gravity constraints on our model. Finally, in Section VI, we present our conclusions and discussion.

\section{Model And Cosmological Implication}
\subsection{Model}
The action of modified $f(R)$ gravity with matter is
\beq
\label{action}
S=\int d^4x\sqrt{-g}(\frac{f(R)}{2\kappa^2}+L_m),
\eeq
where $g$ is the determinant of the metric and R is the Ricci scalar curvature. Taking the variation of the action (\ref{action}) with respect to $g_{\mu\nu}$, we have the equations of motion
\beq
\label{eom1}
R_{\mu\nu}-\frac{1}{2}R g_{\mu\nu} = \frac{\kappa^2}{f'} T^{m}_{\mu\nu}+\frac{1}{\kappa^2 f'}\{\frac{1}{2}g_{\mu\nu}[f(R)-R f']+(\nabla_{\mu} \nabla_{\nu}-g_{\mu\nu}\Box)f'\}.
\eeq
Here $R_{\mu\nu}$ is the Ricci tensor, $f'=df/dR$, $\nabla_{\mu}$ is the covariant derivative operator associated with the metric $g_{\mu\nu}$, and $\Box\phi\equiv g^{\mu\nu}\nabla_{\mu}\nabla_{\nu}\phi$. And $T^{{\mathrm{m}}}_{\mu\nu}$ is the matter stress-energy tensor which satisfies the continuity equation
\beq
\nabla^{\mu}T^m_{\mu\nu}=0.
\eeq
The trace of Eq. (\ref{eom1}) gives
\beq
\label{eom2}
3\Box f'+f'R-2f=\kappa^2T,
\eeq
where T is the trace of $T^m_{\mu\nu}$.

The Einstein gravity, with a cosmological constant, corresponds to $f(R)=R-2\Lambda$,
$f'=1$ and $R=4\Lambda-\kappa^2T$. Especially, in the matter dominated
epoch, $R\simeq-\kappa^2T$. In a general modified gravity, $f'$ could be considered as a new scalar degree of
freedom, $\phi\propto f'$. The trace equation (\ref{eom2}) determines the dynamics of this scalar field
$\phi$:
\beq
\label{eom3}
\Box f'=\frac{\partial V_{eff}}{\partial f'},
\eeq
with the effective potential
\beq
\label{veff}
\frac{\pa V_{eff}}{\pa f'}=\frac{1}{3}(f'R-2f+\kappa^2 \rho).
\eeq
In the late-time universe, $\rho \ll 1$ can be neglected, the equation $\pa V_{eff}/\pa f'=0$
and the stability condition $\pa^2 V_{eff}/\pa f'^2>0$ gives
\beq
\label{dec}
2f(R)-R f'(R)=0
\eeq
and
\beq
\label{stablec}
\frac{\pa^2 V_{eff}}{\pa f'^2}=\frac{1}{3}(\frac{f'}{f''}-R)>0.
\eeq
 If in a $f(R)$ gravity the solution of Eq.(\ref{dec}) gives a positive scalar curvature and satisfies the stability condition (\ref{stablec}), then the universe will enter into a stable de-Sitter phase in the asymptotic future.

To be  cosmologically viable, a $f(R)$ gravity model should satisfy a few requirements:
\begin{itemize}
\item In the high red-shift regime, $f(R)\rightarrow R-2\Lambda$, and mimic the $\Lambda$CDM model which is well tested by the CMB, supernova and other experiments;
\item In the asymptotic future, the model will have a stable vacuum, which is usually a de Sitter spacetime;
\item It should satisfy the local gravity constraints in the high curvature region.
\end{itemize}
  In the cosmologically viable $f(R)$ models in the literature, they usually take a form as $R+g(R/R_0)$, where
$g(R/R_0)$ tends to be a constant when $R/R_0\rightarrow \infty$. We select two well-studied ones\cite{hu,linder} to compare with our model:\\
{\begin{tabular}{@{}ccccc@{}}
&& Model & \hs{6ex}$f(R)$ &\hs{6ex} Parameters \\
&&(i) Hu-Sawicki &   \hs{6ex}  $ R - \frac{c_1 R_{\mathrm{HS}} \left(R/R_{\mathrm{HS}}\right)^p}{c_2
\left(R/R_{\mathrm{HS}}\right)^p + 1}$ &\hs{6ex}$c_1$, $c_2$, $p(>0)$,$R_{\mathrm{HS}}(>0)$\\
&&(ii) Exponential &   \hs{6ex}  $R -\beta R_{\mathrm{E}}\left(1-e^{-R/R_{\mathrm{E}}}
\right)$ & \hs{6ex}$\beta$, $R_{\mathrm{E}}$\\
&&(iii) Our model &   \hs{6ex}  $(R-\lambda c)e^{\lambda(\frac{c}{R})^n}$ & \hs{6ex}$\lambda\geq1$, $n>3$, $c>0$\\
\end{tabular} \label{Table}}

In the matter-dominated epoch, the curvature is large and the exponential factor in our action could be safely set to $1$ and the model reduces to $\Lambda$CDM model with $\Lambda=\frac{\lambda c}{2}$.

To study the asymptotic behavior of our model, we need to solve Eq.(\ref{dec}), taking into account of the stability condition (\ref{stablec}). From the explicit form (\ref{f}),
we calculate the first and the second derivative of $f$
\bea
\label{1st}
f'&=&\left[1-\lambda n(\frac{c}{R})^n+\lambda^2n(\frac{c}{R})^{n+1}\right]e^{\lambda(\frac{c}{R})^n},\\
\label{2cd}
f''&=&\left[\lambda(n^2-n)\frac{c^n}{R^{n+1}}-\lambda^2(n^2+n)\frac{c^{n+1}}{R^{n+2}}
+\lambda^2 n^2\frac{c^{2n}}{R^{2n+1}}-\lambda^3 n^2\frac{c^{2n+1}}{R^{2n+2}}\right]e^{\lambda(\frac{c}{R})^n}.
\eea
If $\lambda$ is large enough or $n\gg 3$, the de-Sitter solution of Eq.(\ref{dec}) is
\beq
R_{ds}\simeq 2\lambda c,
\eeq
which is the same as the solution in the $\Lambda$CDM. For general values of $\lambda$ and $n$, $R_{ds}\neq 2\lambda c$. We will discuss this issue via numerical analysis in the next section.

On the other hand, the stability condition
\bea
\begin{aligned}
\frac{\pa^2 V_{eff}}{\pa f'^2}&=\frac{1}{3}R_{ds}(\frac{f'}{f''R_{ds}}-1)\\
                              &\simeq\frac{1}{3}R_{ds}\left[\frac{R^n_{ds}}{\lambda nc^n[n(1-\frac{\lambda c}{R_{ds}})-(1+\frac{\lambda c}{R_{ds}})]}-1\right]
\end{aligned}
\eea
is easily satisfied when $n>(1+\frac{\lambda c}{R_{ds}})/(1-\frac{\lambda c}{R_{ds}})\simeq3$. If we define
\beq
m\equiv\frac{f''(R)R}{f'(R)},
\eeq
we find that the stability condition becomes simply $0<m<1$, noticing that the Compton wavelength of the extra scalar mode is defined as
\bea
\begin{aligned}
\lambda_c&=(\frac{\pa V_{eff}}{\pa f'})^{-\frac{1}{2}}\\
&\simeq\sqrt{3f''}\simeq\sqrt{3m/R}.
\end{aligned}
\eea
In order to analyze the stability precisely, we have plotted in Fig. 1
the value of parameter $m$ in the asymptotic de-Sitter phase for different values of $n$. When $n>3$, $0<m(R_{ds})<1$ such that the
stability condition is well satisfied. However, when $n\leq3$, the model is unstable. Another feature learned from the Fig.1 is that when $\lambda\geq 4$, the value of $m$ is very tiny.
\begin{figure}
\psfig{figure=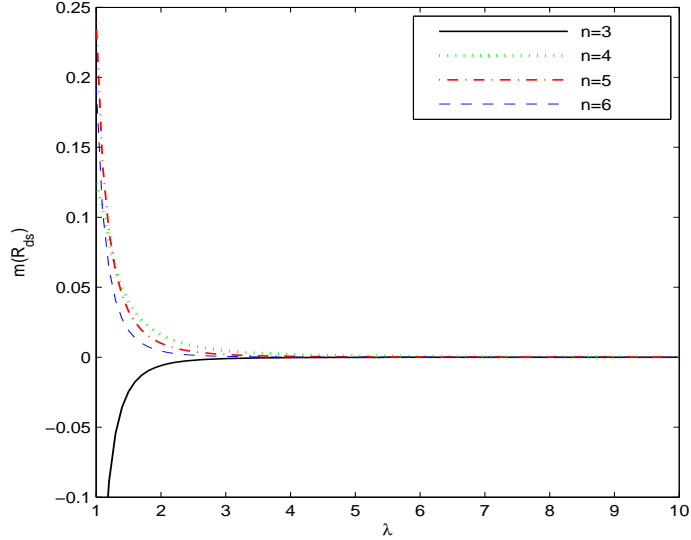,width=10cm,height=8cm}
\caption{The parameter $m(R_{ds})$ as a function of $\lambda$ with $n=3,4,5,6$ at the de-Sitter universe.}
\end{figure}

We will work in the FRW space-time. In this case, Eq. (\ref{eom1}) gives the modified Friedman equations:
\bea
\label{friedman1}
3f'H^2&=&\kappa^2\rho_m+\frac{1}{2}(f'R-f)-3H\dot{f'},\\
\label{friedman2}
-2f'\dot{H}&=&\kappa^2(\rho_m+P_m)+\ddot{f}-H\dot{f},
\eea
in which we have neglected the radiation component because we only focus on the cosmology from matter-dominated epoch to the asymptotic future in this paper. The dot denotes time derivative $\pa/\pa t$. The matter density satisfies the conservation
law
\beq
\dot{\rho}_m+3H(\rho_m)=0,
\eeq
and can be written as $\rho_m=\rho_{0} a^{-3}$. The dark energy density and pressure can be
defined respectively as:
\bea
\label{rho}
\kappa^2\rho_{de}&=&\frac{1}{2}(f'R-f)-3H\dot{f'}+3H^2(1-f'),\\
\label{pressure}
\kappa^2p_{de}&=&\ddot{f'}+2H\dot{f'}-\frac{1}{2}(f'R-f)-(2\dot{H}+3H^2)(1-f').
\eea

\subsection{Dynamical evolution}
According to the work of Linder\cite{linder}, with some modifications, we define
\bea
\label{dynamics}
x_H&=&\frac{H^2}{m^2_0}-a^{-3}-\frac{\lambda c}{6m_0^2},\\
x_R&=&\frac{R}{m^2_0}-3a^{-3}-2\frac{\lambda c}{m_0^2}-12x_H,
\eea
where the parameter $m_0^2$ is defined as $\kappa^2\rho_{0}/3$. The matter density parameter is defined as
\beq
\label{om}
\Omega_m=\frac{a^{-3}}{x_H+a^{-3}+\frac{\lambda c}{6m_0^2}}
\eeq

Using Eq. (\ref{friedman1}), we get
\beq
3f'H^2=\kappa^2\rho_m+\frac{1}{2}(f'R-f)-3H^2f''\frac{dR}{dN},
\eeq
where $N=\ln a$. Using $R=12H^2+3d(H^2)/dN$, the modified Friedman equations become two first-order equations
\bea
\label{dynamics1}
\frac{dx_H}{dN}&=&\frac{x_R}{3},\\
\frac{dx_R}{dN}&=&9a^{-3}-4x_R+\frac{3a^{-3}+\frac{1}{2}(f'R-f)/m_0^2-3f'H^2/m^2_0}{3H^2f''}.
\eea

Here, $R$ and $H$ can be expressed in terms of  the parameters $x_H$ and $x_R$. The dark energy equation of state is defined as
\bea
\label{wde}
\omega_{de}\equiv \frac{-2\dot{H}-3H^2}{3H^2-\kappa^2\rho_m}=-1-\frac{1}{9}\frac{x_R}{x_H+\lambda c/6m^2_0}.
\eea

Now let us study the expansion history of the universe in our model. At the high curvature region dominated by matter, according to the trace equation (\ref{eom2}), the new scalar $f'$ is always near the minimum of the effective
potential,  the Ricci scalar is approximatively. So, we give the expression of Ricci scalar
\bea
R\simeq\kappa^2\rho_m+2\lambda c+\mathcal{O}(R(c/R)^n).
\eea
On the other hand, $f(R)\rightarrow R-\lambda c$ mimic $R-2\Lambda$. Using Eq. (\ref{friedman1}), and
taking the initial value of $\dot{f}'$ as zero, the Hubble parameter can be written as
\bea
H^2=\frac{\kappa^2\rho_m}{3}+\frac{\lambda c}{6}+\mathcal{O}(R(c/R)^n).
\eea
 In the early universe, $x_H$ and $x_R$ are the same order as $R(c/R)^n$ and can be neglected. Consequently, we may take the initial condition $x_H=x_R=0$ at $z$=13. 

In the asymptotic future, the universe will go into a de-Sitter epoch as expected. The stationary point of the equation of motion
is given as
\bea
&x_R=0,\\
&\frac{1}{2}(f'R-f)-3f'H^2=0,
\eea
which is equivalent to Eq. (\ref{dec}). The fact that $x_{R}$ evolves from zero to zero suggests that the equation of state of dark energy starts from $-1$ at the matter-dominated epoch
and will end with the same value  $-1$  in the future. With the matter density becoming smaller and smaller, the Ricci scalar would tend to be a constant $R_{ds}$.

\subsection{Numerical analysis}

\begin{figure}
\psfig{figure=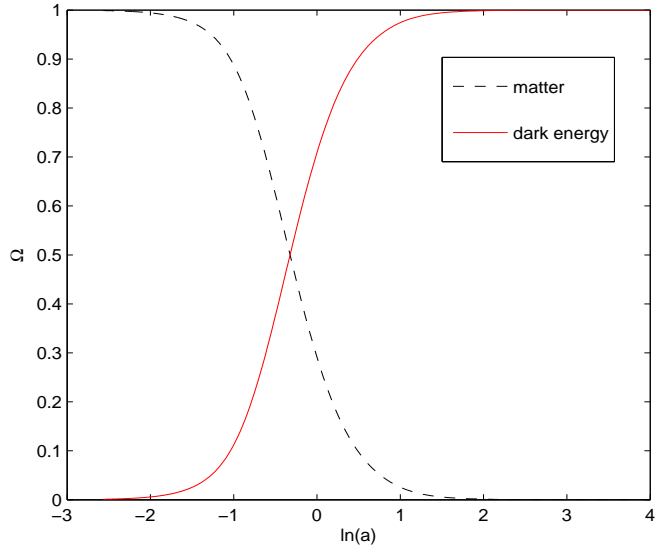,width=10cm,height=8cm}
\caption{Density parameters $\Omega_m$ and $\Omega_{\Lambda}$ in our model with the parameters
$\lambda=1$, n=4.}
\end{figure}

\begin{figure}
\psfig{figure=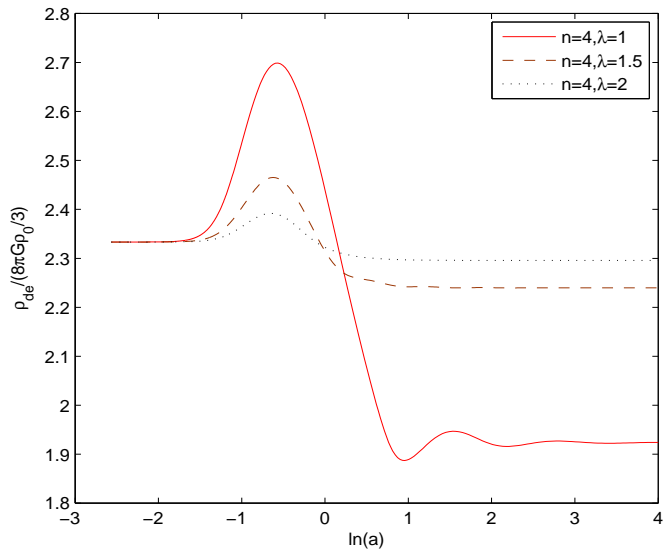,width=10cm,height=8cm}
\caption{Dark energy density $\rho_{de}$, normalized by $\kappa^2\rho_0/3$.}
\end{figure}

\begin{figure}
\psfig{figure=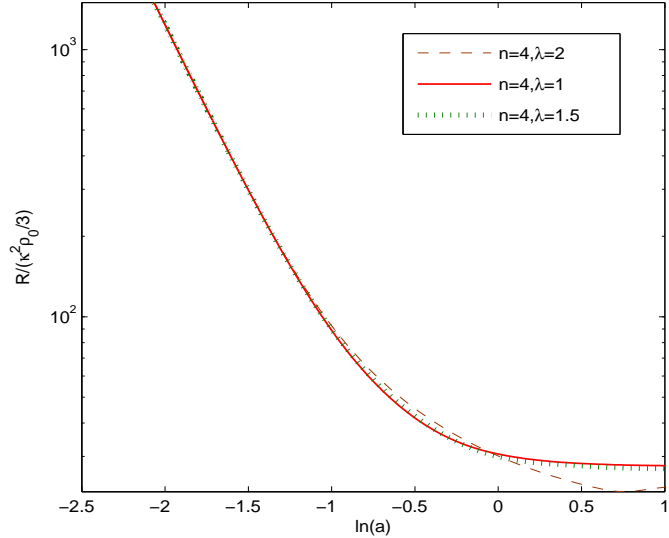,width=10cm,height=8cm}
\caption{The Ricci scalar curvature, normalized by $\kappa^2\rho_0/3$ evolves from high curvature to a constant.}
\end{figure}

\begin{figure}
\psfig{figure=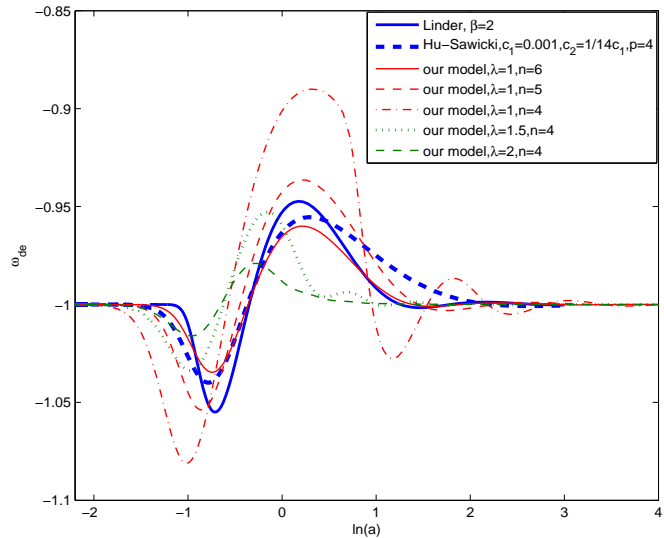,width=10cm,height=8cm}
\caption{The dark energy equation of state of for different models.}
\end{figure}

\begin{figure}
\psfig{figure=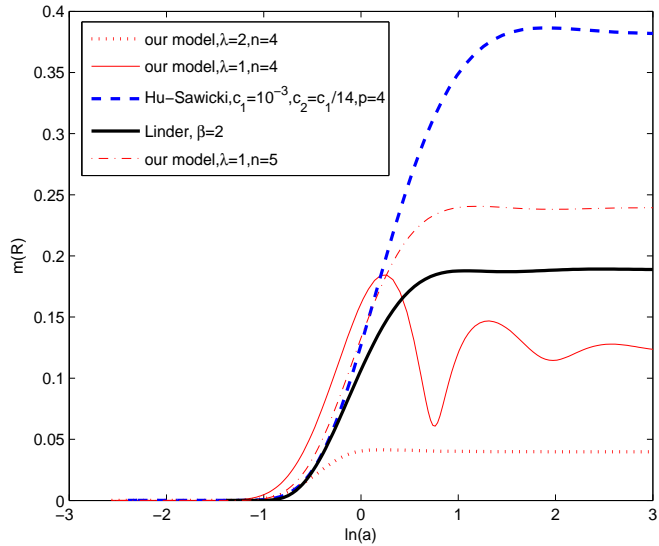,width=10cm,height=8cm}
\caption{The parameter $m$ as a function of $\ln$$a$.}
\label{fig-m}
\end{figure}
\begin{figure}
\psfig{figure=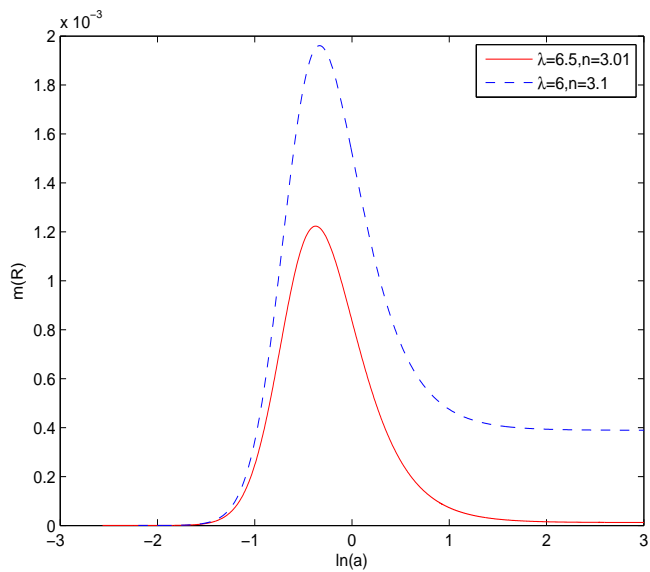,width=10cm,height=8cm}
\caption{The parameter $m$ as a function of $\ln$$a$ with parameter $n\simeq3$.}
\label{fig-m}
\end{figure}

\begin{figure}
\psfig{figure=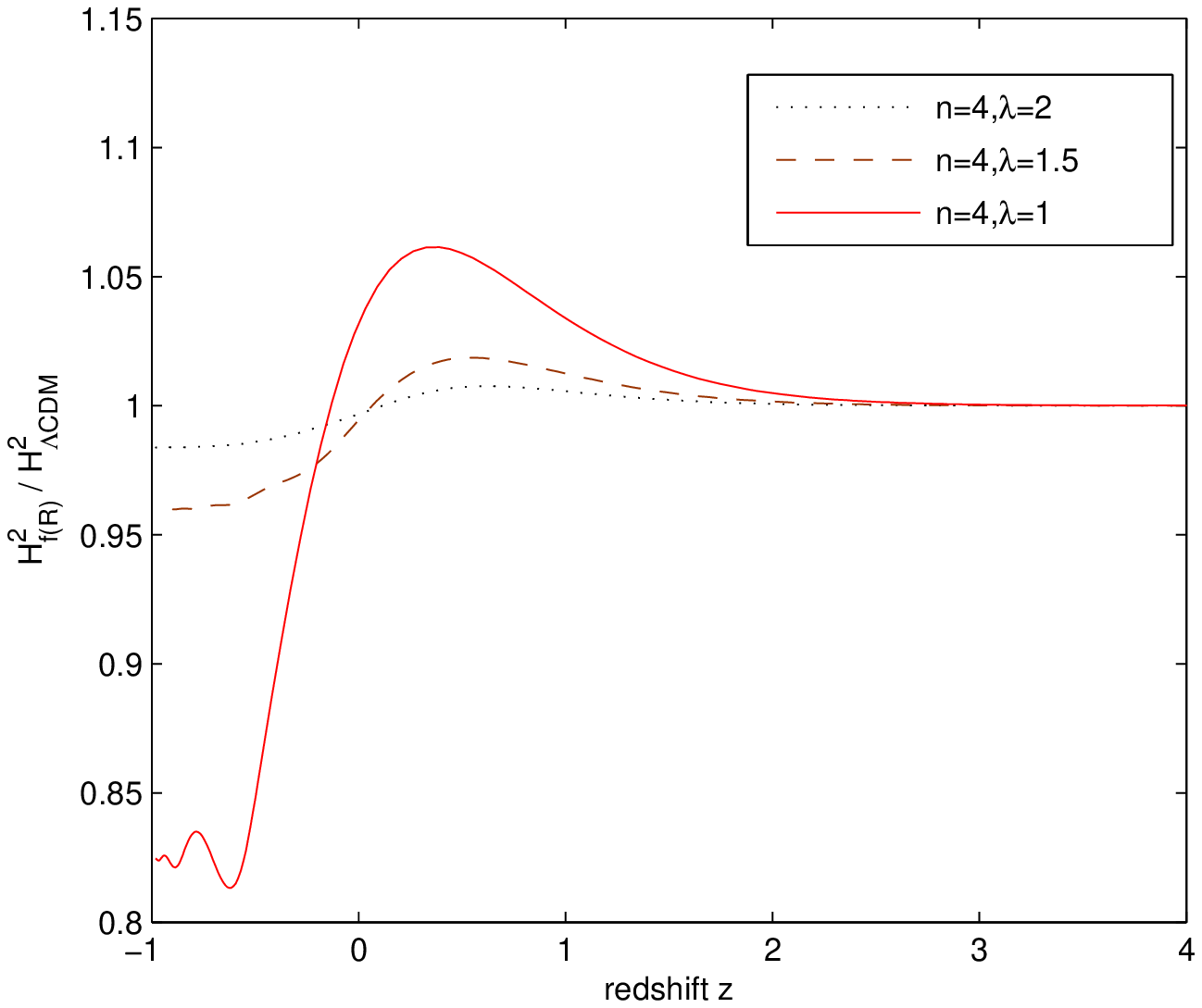,width=10cm,height=8cm}
\caption{The square of Hubble parameter in our model comparing with $\Lambda CDM$, and its evolution.}
\end{figure}

From Eq. (\ref{om}), we plot the density parameters $\Omega_m$ and $\Omega_{de}$ as a
function of ln$a$ in Fig. 2. Note that, the current matter density parameter $\Omega_m\simeq0.3$ is in accordance with the observation.

In Fig. 3 we show the evolution of the dark energy density in our model. In the high red-shift region $N<-1$, the dark energy density stays around the value $\lambda c/2$. In the region $-1<N<1$, it first rise up to a maximum, and then falls down and oscillates around another constant value asymptotically. Note that the dark energy density tends asymptotically to different values with different parameters $\lambda$. A smaller $\lambda$ leads to a smaller asymptotic value of  dark energy density in the future. When $\lambda$ is large enough, the $\kappa^2\rho_{de}$ would tend to be $\lambda c/2$, as the same as the value in the matter-dominated epoch.

In Fig. 4, we show the evolution of the Ricci scalar curvature for different values of $\lambda$. We see that $R$ evolves
from a high curvature and quickly reaches its asymptotic value $R_{ds}$. Note that a smaller $\lambda$ will lead
to a smaller value of $R_{ds}$.

In Fig. 5, we show the evolution of the equation of state of dark energy for different models and parameters. From the dark energy
conservation equation, we know $\omega_{de}=-1-\dot{\rho}_{de}/(3H\rho_{de})$. We can see the relationship between $\omega_{de}$ and $\rho_{de}$ from Fig. 3 and Fig. 5. At the beginning, $\omega_{de}$ stays near $ -1$ in high red-shift region $N<-1$. It first falls down to a minimum  and  then climbs up to a maximum. It falls down again and begin oscillating  around $-1$ with smaller amplitude. It finally settle down to $-1$. We compare the models of Linder and Hu-Saweicki with ours. We find that in all models, the evolution of dark energy equation of state is quite similar qualitatively. However, the details of the evolution are different. In particular, when $n\leq 4$, the oscillation behavior of our model looks different. For example,
 in the case $n=4$, in our model  $\omega_{de}$ cross phantom divide line $-1$ earlier and oscillates in the region $-1<\ln a<1$. Fig. 5 indicates also that a larger value of the parameter $n$ will reduce the deviation from $-1$, seeing the red lines with parameters $n=4, 5, 6, \lambda=1$.  So does a larger value of $\lambda$,  seeing the lines with parameters $\lambda=2, 1.5, 1, n=4$. The fact that  larger values of $n$ and $\lambda$ will suppress the deviation of $\omega_{de}$ from $-1$ is easy to understand: the larger $n$ or $\lambda$, the closer our model to the $\Lambda$CDM model.


In Fig. 6 and Fig. 7, we show the evolution of the parameter $m$ which decides the Compton wavelength $\lambda_c=\sqrt{3m/R}$. The parameter $m$ in the models of Hu-Saweicki and Linder starts from a small quantity, then climbs up to a constant. In our model, when $n\leq 4$, the oscillation of the parameter $m$ in the evolution is obvious. This difference accounts for the fact that the dark energy equation of state in our model has a different oscillating form from the ones in the other two models. When $n=5, 6$ the shape of the evolution lines is similar to the ones in the other two models. In general, a larger $\lambda$ will lead to a smaller $m$. When $n\rightarrow3$, the terms of $(c/R)^n$ and $(c/R)^{n+1}$ in the expression of $f''$ cancel out each other, the terms of $(c/R)^{2n}$ and $(c/R)^{2n+1}$ become important. Therefore the parameter $m$ becomes even smaller asymptotically, seeing Fig. 7.

In Fig. 8, we show the evolution of the Hubble parameter in our model, comparing with the $\Lambda$CDM model. We see, in the high red-shift region, the ratio tends to be $1$, which means the $f(R)$ is close to the $\Lambda$CDM. However in the low red-shift region, it will be a little larger than that in $\Lambda$CDM. And in the future $-1<z<0$,
the ratio will be less than $1$.  Note that, a smaller value of the parameter
$\lambda$ suggests  a larger deviation from $\Lambda$CDM in the late-time universe. As a result, the current age of the universe in our model is smaller than the one in the $\Lambda$CDM model. Moreover, the cosmological distance is also be affected. Thus we can use the observational data to constrain the parameters. For $n=4$, the deviation from the $\Lambda$CDM model appears at $z=3$.

\section{The local gravity test}

In this section, we discuss the compatibility of our model with the local gravity test. As a $f(R)$ gravity theory is equivalent to a scalar-tensor theory,  the extra scalar mode may mediate a long range attractive fifth force and  thus violate solar system constraints. However, it has been suggested in \cite{hu,tsu1,tsu2,tf,david,chameleon,Khoury:2003aq}, the $f(R)$ models can be consistent with the local gravity constraints with the help of the chameleon mechanism.
In the chameleon mechanism, the scalar behaves differently in different environments. This
is achievable in $f(R)$ gravity as the scalar potential get modified by the scalar coupling to the matter density.

It is more convenient to work in the Einstein frame, which is related to the original Jordan frame by a Weyl scaling $\widetilde{g}_{\mu\nu}=f'g_{\mu\nu}$. The extra scalar field $\phi$ is defined as $2\beta\kappa\phi=\ln f'$ with $\beta=1/\sqrt{6}$. The action in the Einstein frame is
\beq
S_{EF}=\int
\label{aef}
\sqrt{-\widetilde{g}}(\frac{m_{pl}^2}{2}\widetilde{R}-\frac{1}{2}\widetilde{g}_{\mu\nu}\partial_{\mu}\phi\partial_{\nu}\phi-V(\phi))
+\int d^4XL_{m}(\widetilde{g}_{\mu\nu}f'^{-1},\psi_{M}),
\eeq
where
\bea
V(\phi)=\frac{f'R-f}{2\kappa^2f'^2}
\eea
If we consider the background geometry as a Minkowski spacetime, then the equation of motion for the scalar $\phi$ in the Einstein frame is just
\bea
\label{eomphi}
\frac{d^2\phi}{d\widetilde{r}^2}+\frac{2}{\widetilde{r}}\frac{d\phi}{d\widetilde{r}}-\frac{dV_{eff}}{d\phi}=0
\eea
where $\widetilde{r}$ is the distance from the center of the object, and $V_{eff}$ is the effective potential,
\bea
V_{eff}=V(\phi)+\rho_m e^{-\kappa\beta\phi}.
\eea
It is this form of the potential that makes the chameleon mechanism feasible.

 Let us consider a spherically symmetric object with a radius $\widetilde{r}_c$, mass $M_c$, a matter density $\rho=\rho_A$ when $\widetilde{r}<\widetilde{r}_c$,  and
$\rho=\rho_B$ when $\widetilde{r}>\widetilde{r}_c$. The effective potential has a different shape with a different surrounding matter density. It has minima at $\phi_A$ and $\phi_B$:
\bea
V,_{\phi}(\phi_A)-\rho_A\beta\kappa e^{-\beta\kappa\phi_A}=0,\\
V,_{\phi}(\phi_B)-\rho_B\beta\kappa e^{-\beta\kappa\phi_B}=0,
\eea
where the scalar has different mass $m^2_A=\pa^2 V_{eff}(\phi_A)/\pa\phi^2$ and $m^2_B=\pa^2 V_{eff}(\phi_B)/\pa\phi^2$ respectively.
For our model, the first derivative of the effective potential is
\bea
\label{phimin}
\begin{aligned}
\frac{dV_{eff}}{d\phi}&=\frac{\pa V}{\pa R}\frac{\pa R}{\pa \phi} -\beta\kappa\rho_m e^{-\beta\kappa\phi}\\
                      &=\beta\kappa\left[\frac{(2f-f'R)}{\kappa^2f'^3}-\rho_m e^{-\beta\kappa\phi}\right]\\
                      &\simeq\beta\kappa\left(\frac{2f-f'R}{\kappa^2}-\rho_m\right),
\end{aligned}
\eea
where we have used $f'=e^{2\beta\kappa\phi}\simeq1$ and $f''=2\beta\kappa e^{2\beta\kappa\phi}\frac{\pa\phi}{\pa R}$ because that the minimum of the potential is very near $0$ in the high curvature region $R_0\simeq\kappa^2\rho_m\gg1$. According to the definition we have,
\bea
2\beta\kappa\phi_{min}=\ln f'\simeq-\lambda(n-1)(\frac{c}{R_0})^n+\lambda^2n(\frac{c}{R_0})^{n+1}.
\eea
The second derivative of the potential gives the scalar mass
\bea
\label{mass}
M^2(\phi)=\frac{d^2V_{eff}}{d\phi^2}=\frac{R_0}{3}(\frac{1}{m(R_0)}-\frac{7}{2}).
\eea
If the object satisfies the thin-shell condition $\kappa(\phi_B-\phi_A)/6\beta\Phi_c\ll1$, where $\Phi_c$ is the Newton potential,
then the scalar has an exterior solution
\bea
\label{phir}
\phi(r)=\phi_B-\frac{\phi_B-\phi_A}{\Phi_c}\frac{GM_c}{\widetilde{r}}e^{-M_B(\widetilde{r}-\widetilde{r_c})}.
\eea
If the object is the Sun,  we have $\rho_A\simeq1g/cm^3$ and $\rho_B=10^{-24}g/cm^3$ which is the galaxy matter density. The cosmological density is $\rho_0=10^{-29}g/cm^3$. Thus $R_A\simeq\kappa^2\rho_A$ is much larger than $R_B$, and according to Eq. (\ref{phimin}), we have
\bea
\begin{aligned}
2\beta\kappa\phi_B &\simeq -\lambda^{-n+1}(n-1)(\frac{\lambda c}{\kappa^2\rho_B})^n\\
                    &\simeq -\lambda^{-n+1}(n-1)(\frac{\lambda c}{\kappa^2\rho_0})^n10^{-5n}\\
                    &\simeq -\lambda^{-n+1}(n-1)(\frac{2\Omega_{\Lambda}}{\Omega_m})^n10^{-5n},
\end{aligned}
\eea
where we have used the fact that $\lambda c\rightarrow 2\Lambda$. Obviously $\phi_B$ is much larger than
$\phi_A$ which can be neglected in Eq. (\ref{phir}). From Eq. (\ref{mass}), we have
\bea
\begin{aligned}
M^2_B   &\simeq   \frac{\kappa^2\rho_B}{3m(R_B)}\\
        &\simeq   10^{-24}\frac{\kappa^2\rho_A}{3}\left(\frac{2\Omega_{\Lambda}}{\Omega_m}\right)^n\\
        &\simeq   \frac{10^{-18}}{\widetilde{r}^2_{c}},
\end{aligned}
\eea
where we have used $\Phi_c=GM_c/\widetilde{r}_c=\kappa^2\rho_A\widetilde{r}^2_c/6\simeq10^{-6}$. We see that the mass of the scalar is very light
out of the object, so the exponential $e^{-M_B(\widetilde{r}-\widetilde{r_c})}$ in Eq. (\ref{phir}) could be set to $1$.

 Let us transform back to the Jordan frame. Under the inverse transformation, $g_{\mu\nu}=e^{-2\beta\kappa\phi}\widetilde{g}_{\mu\nu}$, $\widetilde{r}=e^{2\beta\kappa\phi}r$, the metric in the Jordan frame is
\bea
ds^2=e^{-2\beta\kappa\phi}d\widetilde{s}^2=-[1-2\mathcal{A}(r)]dt^2+[1+2\mathcal{B}]dr^2+r^2d\Omega^2.
\eea
under the condition $\beta\kappa\phi\ll1$.  Then we have the following relations
\bea
\mathcal{A}(r)&\simeq&\widetilde{\mathcal{A}}(\widetilde{r})+\beta\kappa\phi\
\simeq\frac{GM_c}{r}[1+\frac{\kappa\beta\phi_B}{\Phi_c}(\frac{r}{r_c}-1)],\\
\mathcal{B}(r)&\simeq&\widetilde{\mathcal{B}}(\widetilde{r})+\beta\kappa\widetilde{r}\frac{d\phi}{d\widetilde{r}}
\simeq\frac{GM_c}{r}(1+\beta\kappa\frac{\phi_B}{\Phi_c}).
\eea
 The tightest experimental bound on the PNP parameters is given by $|\gamma-1|<2.3\times10^{-5}$\cite{tsu3,616,617,bb}. If we take the distance $r=r_c$, then we can constrain the parameter in our model
 \bea
 |\beta\kappa\phi_B|=\frac{1}{2}\lambda^{-n+1}(n-1)(\frac{2\Omega_{\Lambda}}{\Omega_m})^n10^{-5n}
                     <2.3\times10^{-11}.
 \eea
As $n > 3$ and  $\lambda\geq1$ in our model, this bound can be satisfied easily.

On the other hand, the experiment from the violation of equivalence principle gives a slightly more stringent bound\cite{chameleon,tsu3,tsu2,616},
\bea
|\frac{\kappa\phi_B}{6\beta\Phi_c}|<8.8\times10^{-7}/\beta,
\eea
which can be translated into
\bea
|\beta\kappa\phi_B|<2.1\times10^{-15}.
\eea
It could be evaded in our model without trouble.

\section{Conclusions and discussion}

In this paper, we have proposed a new viable $f(R)$ dark energy model. It is of an exponential form, but is different from the Exponential gravity proposed in \cite{linder}. We focus on the cosmological evolution starting from matter-dominated epoch to the asymptotic future. In the matter-dominated epoch, our model reduces to the $\Lambda$CDM model with a positive cosmological constant. In the asymptotic future, the universe will settle down to a stable de-Sitter phase.  However, between two epoches, the evolution of the universe in our model could be very different from the $\Lambda$CDM model, if we choose the parameters appropriately.

The solar system constraints of $f(R)$ gravity place weak bounds on our model. Due to the exponential suppression, our model has little deviation from GR in the high curvature region. It can evade local gravity constraints easily.

It turns out that in the parameter range $3< n\leq 4, \lambda \sim 1$, the model is most distinguishable from other models. In our model, the dark energy equation of state cross the phantom divide line many times, before it asymptotically settle down to a constant $-1$. Comparing with the other $f(R)$ models, the dark energy equation of state in our model has  strong oscillations, and will cross the phantom divide line  in the earlier future. This prediction can be tested in the future observation. In our model, the Hubble parameter shows the deviation from the $\Lambda$CDM model at the redshift $z\approx 3$, suggesting that the current age of the universe and the theoretical distance of the cosmological objects may be smaller in our model. It would be very interesting to further constrain the parameters in our model from cosmological observations.


\noindent
\section*{Acknowledgments}

The work was in part supported by NSFC Grant No. 10975005.

\end{document}